%% file: main.tex
\documentclass{llncs}
\usepackage{hyperref}
\usepackage{xspace}
\usepackage{alltt}
\usepackage[T1]{fontenc}
\usepackage[subtle,mathspacing=normal]{savetrees}

\newcommand{\runbibtex}[1]{}
\newcommand{\usebbl}[1]{#1}

\newcommand{\fullonly}[1]{}

\newcommand{\lncsonly}[1]{#1}
\newcommand{\articleonly}[1]{}
\newcommand{\acmonly}[1]{}
\newcommand{\svonly}[1]{}

\newcommand{\mycomment}[1]{}

\acmonly{
\copyrightyear{2017} 
\acmYear{2017} 
\setcopyright{acmcopyright}
\acmConference{SACMAT '17}{June 21--23, 2017}{Indianapolis, IN, USA}
\acmPrice{15.00}
\acmDOI{http://dx.doi.org/10.1145/3078861.3078878}
\acmISBN{978-1-4503-4702-0/17/06}
}

\articleonly{}
\lncsonly{}
\acmonly{}

\newcommand{\eg}[1]{}

\newcommand{\Hex}[1]{\hspace{#1ex}}
\newcommand{\Vex}[1]{\vspace{#1ex}}
\newenvironment{code}{\Vex{-.0}\begin{alltt}\small}{\end{alltt}\Vex{-.0}}
\newcommand\co[1]{\mbox{\small\tt #1}} 
\newcommand\Sp[1]{\Hex{0}}
\newcommand{\mathify}[1]{\ifmmode{\mbox{$#1$}}\else\mbox{$#1$}\fi}
\newcommand\p[1]{{\small \mathify{\it #1}}} 

\begin{document}

\newcommand{\thanksText}{This material is based on work supported in part by %
    ONR Grant N00014-15-1-2208, 
    NSF Grants %
    CCF-1414078 
    and CNS-1421893, 
    and DARPA Contract FA8650-15-C-7561. 
}

\title{Algorithm Diversity for Resilient Systems}

\articleonly{\author{Scott~D.~Stoller \and Yanhong~A.~Liu\\
    Department of Computer Science, Stony Brook University, USA}}

\lncsonly{\author{Scott~D.~Stoller \and Yanhong~A.~Liu}
  \institute{Department of Computer Science, Stony Brook University, USA\\ \email{\{stoller,liu\}@cs.stonybrook.edu}}}

\acmonly{
\author{Scott~D.~Stoller}
\affiliation{
\institution{Stony Brook University}}
\email{stoller@cs.stonybrook.edu}
\author{Yanhong~A.~Liu}
\affiliation{
\institution{Stony Brook University}}
\email{liu@cs.stonybrook.edu}
}

\svonly{\author{Scott~D.~Stoller \and Yanhong~A.~Liu}

\institute{S. D. Stoller \and Yanhong~A.~Liu \at Stony Brook University, USA\\ \email{stoller@cs.stonybrook.edu}}

\date{Received: date / Accepted: date}}

\newcommand{\abstracttext}{ %
Diversity can significantly increase the resilience of systems, by reducing the prevalence of shared vulnerabilities and making vulnerabilities harder to exploit.   Work on software diversity for security typically creates variants of a program using low-level code transformations.  This paper is the first to study {\em algorithm diversity} for resilience.  We first describe how a method based on high-level invariants and systematic incrementalization can be used to create algorithm variants.  Executing multiple variants in parallel and comparing their outputs provides greater resilience than executing one variant.  To prevent different parallel schedules from causing variants' behaviors to diverge, we present a {\em synchronized execution} algorithm for DistAlgo, an extension of Python for high-level, precise, executable specifications of distributed algorithms.  We propose static and dynamic metrics for measuring diversity.  An experimental evaluation of algorithm diversity combined with implementation-level diversity for several sequential algorithms and distributed algorithms shows the benefits of algorithm diversity.
}

\acmonly{
\begin{abstract}
\abstracttext
\end{abstract}}




\acmonly{\thanks{\thanksText}}

\maketitle

\lncsonly{
\begin{abstract}
\abstracttext
\end{abstract}}
\articleonly{
\begin{abstract}
\abstracttext
\end{abstract}}
\svonly{
\begin{abstract}
\abstracttext
\end{abstract}}

\input{intro.tex}
\input{distalgo.tex}

\input{inc.tex}
\input{sync.tex}

\input{metrics.tex}
\input{evaluation.tex}
\input{related.tex}

\paragraph{Acknowledgements.}

\thanksText %
We thank Thang Bui, Rahul Gadi, Shikhar Sharma, Shalaka Sidmul, Shubham Singhal, and Swetha Tatavarthy for their contributions to the implementation and experiments.
  


%
\usebbl{\input{main.bbl}}
\runbibtex{
\acmonly{\bibliographystyle{ACM-Reference-Format}}\lncsonly{\bibliographystyle{splncs04}}\articleonly{\bibliographystyle{alpha}}\svonly{\bibliographystyle{plain}}
\bibliography{diversity,../liubib/strings,../liubib/liu,../liubib/PT,../liubib/Sys,../liubib/crossref}}

\end{document}

%% file: intro.tex

\section{Introduction}
\label{sec:intro}

Diversity can significantly increase the resilience of systems, by reducing the prevalence of shared vulnerabilities and making vulnerabilities harder to exploit.  The idea of intentionally introducing software diversity as a defense mechanism has been around for decades, e.g., \cite{cohen93evolution,forrest1997building}.  It is closely related to the well-known {\em moving target defense} (MTD) strategy: running different variants of a program at different times is MTD.  Software diversity is an effective defense against attacks whose success depends on details of the victim software.  Without knowing those details for the specific instance (variant) of the software being attacked, attackers can still attempt such attacks (e.g., by making random guesses at those details), but the probability of success is greatly reduced \cite{larsen2014}.

There is a large corpus of research on techniques for automatically introducing software diversity that increase resilience to various classes of attacks \cite{larsen2014}.  For example, Address Space Layout Randomization (ASLR), which randomizes the starting addresses of segments in a process's address space, is a classic form of software diversity that increases resilience to some types of memory corruption attacks. 

The most common way to use software diversity to increase resilience is to run a randomly selected variant each time the program is executed.  With this approach, the use of diversity alters, with high probability, the effect of an attack, so the attack does not have the intended effect (e.g., gaining root privilege and installing a rootkit) \cite{larsen2014}.  The attack might still have a less malicious and less predictable but nevertheless undesirable effect (e.g., crash or incorrect output).


Another way is to run multiple variants of the application in parallel and compare their outputs.  We call this {\em diversified replication}.  Any difference in the outputs of the variants indicates misbehavior of one or more variants due to an attack; this triggers defensive action.  This approach provides greater resilience, at the cost of more computational resources.  It also provides greater resilience than traditional replication, in which replicas are identical and exhibit the same (incorrect) behavior when their vulnerabilities are exploited.  Note that diversity may lead to different behavior (and therefore attack detection) in two ways: (1) it might cause a difference in the direct effect of the attack (e.g., which data structure is overwritten) or, (2) even if the direct effect of the attack is the same (e.g., the same data structure is overwritten), it might cause differences in subsequent behavior, due to differences in the algorithms or implementations used by the variants (e.g., one variant reads the affected data structure earlier in its computation and hence before the attack, and another reads the affected data structure later in its computation and hence after the attack).


This paper focuses on {\em algorithm diversity} for software resilience, in which different variants run different algorithms, i.e., perform different computations at a high level.  In contrast, all of the work surveyed in \cite{larsen2014} creates {\em imple\-men\-ta\-tion-level diversity}, changing details of the implementation without changing the algorithm.  Algorithm diversity can introduce new and larger differences between variants than implementation-level diversity and hence can provide greater resilience, especially when used together with implementation-level diversity. 


Algorithm variants may be obtained in a variety of ways, besides writing them manually.  For standard problems (e.g., dictionary ADT), they can be obtained from algorithm libraries.  A more general automated approach is to generate them by starting with a 
high-level algorithm (or specification) and applying different optimizations (algorithm improvements, automated using program analysis and transformation).  In particular, we have used a method based on systematic incrementalization~\cite{PaiKoe82,Liu+05OptOOP-OOPSLA,Liu+09Inv-GPCE,Liu+16IncOQ-PPDP}, which transforms programs 
to maintain high-level invariants 
incrementally, and related optimizations to generate multiple variants of many sequential algorithms and distributed algorithms~\cite{Liu13book,Liu+12DistPL-OOPSLA,Liu+17DistPL-TOPLAS}.

Algorithm diversity and implementation-level diversity introduce different kinds of variation and together offer more diversity than either alone.  We introduce {\em diversity metrics} that quantify the difference between---or equivalently, the similarity of---variants.   We consider a static metric, {\em code diversity}, based on the instruction sequences in the compiled program, and two dynamic (behavioral) metrics: {\em trace diversity}, based on the sequence of instructions executed, and {\em input access diversity}, based on the sequence of accesses to input data.  The latter dynamic metric is motivated by the fact that invalid inputs are the primary attack vector for external attackers.  A direction for future work is to augment these broad diversity metrics with more specialized metrics that quantify resilience to specific classes of attacks.

Algorithm diversity can be applied to programs in any language.  In this paper, we focus on Python and DistAlgo~\cite{Liu+12DistPL-OOPSLA,Liu+17DistPL-TOPLAS}, an extension of Python for high-level, precise, executable specifications of distributed algorithms.  In contrast, existing work on automated software diversity primarily targets C programs or (disassembled) machine code.  

Python is interpreted---more precisely, CPython, the predominant implementation of Python, compiles Python to bytecode and then runs the bytecode in an interpreter.  Algorithm diversity applied to Python programs can be used together with implementation-level diversity applied to Python programs and the runtime system. 
This achieves greater total diversity and increases resilience to vulnerabilities in the runtime system, because vulnerabilities manifest only with specific inputs, and the runtime system's inputs include Python programs as well as network messages, UI events, etc. 
Diversity at the high-level language level provides additional protection from data-only attacks \cite{noncontrol2005,JITROP2017}, against which many runtime-system-level defenses are less effective.  Algorithm diversity applied to Python programs can also provide resilience to functional faults in the runtime system, if the runtime system does not correctly implement the semantics of some built-in constructs or library functions in some (corner) cases. 


Diversified replication requires {\em synchronized execution} (often called {\em $N$-version execution} \cite{Nversion1985}) of the variants; otherwise, their executions might diverge due to scheduling differences.  Synchronized execution of distributed programs generally requires synchronization of message delivery order.  DistAlgo's asynchronous message handling requires additional synchronization, to ensure that all variants handle corresponding messages at corresponding points in their executions.  We developed a synchronized execution framework for DistAlgo that ensures this.  Our framework can also suppport variants whose behaviors differ in prescribed ways.


Measuring dynamic diversity for Python and DistAlgo programs required development of new runtime monitoring tools, which are also more broadly useful.  We designed and implemented a tool that intercepts accesses to fields of selected objects; we use it to log accesses to objects read as input, including objects received in messages.  Handling built-in types such as integers and strings is tricky, because they are sometimes accessed directly by C code in the Python interpreter, but essential, because they are commonly used in program inputs.

We also designed and implemented a tracing tool that reconstructs the exact sequence of bytecode instructions executed by a Python program.  It uses the standard Python tracing module to record the sequence of source lines executed, and then analyzes the compiled program to determine the sequence of bytecode instructions corresponding to each source line.  Supporting DistAlgo requires some extra work, due to details of DistAlgo's implementation by translation to Python.

In summary, the contributions of this paper include:
\begin{itemize}
\item The first study of semi-automated algorithm diversity for software resilience, using a method based on systematic incrementalization to generate algorithm variants.
\item A synchronized execution framework for DistAlgo and for high-level executable specifications of distributed algorithms.
\item Static and dynamic metrics for measuring diversity.
\item A runtime monitoring tool for Python and DistAlgo that logs accesses to fields of selected objects, including instances of built-in types.
\item A tracing tool for Python and DistAlgo that reconstructs the exact sequence of executed bytecode instructions.
\item Experimental evaluation of algorithm diversity combined with implementation-level diversity for several sequential algorithms and distributed algorithms, demonstrating that algorithm diversity can achieve more diversity than implementation-level diversity, and the two together can achieve even more.
\end{itemize}


%% file: distalgo.tex

\section{Background on DistAlgo}
\label{sec:distalgo}

Liu et al.~\cite{Liu+12DistPL-OOPSLA,Liu+17DistPL-TOPLAS} propose
DistAlgo, a language for high-level, precise, executable specifications of distributed algorithms, and study its use for specification, implementation, optimization, and simplification of such algorithms. For expressing distributed algorithms at a high level, DistAlgo supports four main concepts by building on an object-oriented programming language, Python:~(1) distributed processes that send messages, (2) control flow for handling received messages, (3) high-level queries for synchronization conditions, and (4) configuration for setting up and running.
DistAlgo is specified precisely by a formal operational
semantics~\cite{Liu+17DistPL-TOPLAS}.

\paragraph{Processes That Send Messages.}

A process type $P$ is defined by a class definition for $P$ that inherits from DistAlgo's built-in \co{process} class.  The definition of $P$ may contain, in addition to the usual definitions that may appear in Python classes, definition of a \co{setup} method for taking in and setting up the values used by the process, definition of a \co{run} method containing the main control flow of the process, and definitions of \co{receive} handlers for handling messages, as described below.  

To create instances of $P$, DistAlgo provides a \co{new $P$} construct; it can optionally be preceded by the number of processes to create (the default is 1) and followed by ``\co{at} $h$'' where $h$ identifies the host where the process(es) should be created (the default is the local host).  After a new process has been created, and its \co{setup} method called to initialize it, invoking its \co{start} method causes execution of its \co{run} method.

Processes send messages using the statement \co{send $m$ to $ps$}, where $ps$ is a process or set of processes.

\paragraph{Control Flow for Handling Received Messages.}

Received messages can be handled asynchronously, using \co{receive} definitions, and synchronously, using \co{await} statements. A \co{receive} definition has the form \co{receive $m$ from $p$:~\p{stmt}}.  It handles un-handled messages that match \co{$m$ from $p$}, where $m$ and $p$ are patterns.  If matching succeeds, unbound variables in $m$ (and $p$) are bound to the corresponding component of the message (and the message's sender, respectively), and then \p{stmt} is executed.

To synchronize message handling with local computation, \co{receive} handlers are executed only at {\em yield points}.  The program point before or after any statement can be declared as a yield point.  In addition, there is an implicit yield point before each \co{await} statement, for handling messages while waiting.  By default, any number of pending messages can be handled at a yield point.

An \co{await} statement has the form
\begin{code}
    await \p{cond\sb{1}}:\,\p{stmt\sb{1}} or ... or \p{cond\sb{k}}:\,\p{stmt\sb{k}} timeout\,\p{t}:\,\p{stmt}
\end{code}
It waits until one of \co{\p{cond_1}}, ..., \co{\p{cond_k}} is true or time \co{\p{t}} has elapsed, and then nondeterministically selects one of \co{\p{stmt_1}}, ..., \co{\p{stmt_k}}, \co{\p{stmt}} whose condition is true and executes the selected statement.  Each branch is optional.
 

\paragraph{High-Level Queries for Synchronization Conditions.}

DistAlgo provides constructs to express synchronization conditions in \co{await} statements as high-level queries over message histories (or other sets or sequences).  A query can be an existential or universal quantification, a comprehension, or an aggregation.
An existential quantification has the form \co{some \p{v\sb{1}} in
    \p{s\sb{1}}, ..., \p{v\sb{k}} in \p{s\sb{k}} | \p{cond}}.
It returns true iff \co{\p{cond}} holds for some combination of values of variables that satisfies all \co{\p{v\sb{i}} in \p{s\sb{i}}} clauses.  Universal quantification is similar, with keyword \co{each} instead of \co{some}.

A comprehension has the form
\co{\{\p{e}:~\p{v\sb{1}} in \p{s\sb{1}}, ..., \p{v\sb{k}} in
    \p{s\sb{k}}, \p{cond}\}}. %
It returns the set of values of \co{\p{e}} for all combinations of values of variables that satisfy all \co{\p{v_i} in \p{s\sb{i}}} clauses and condition \co{\p{cond}}.

DistAlgo automatically maintains histories of messages sent and received by each process in variables \co{sent} and \co{received}; they are automatically eliminated if unused.

\paragraph{Configuration.}

Configuration for requirements such as use of logical clocks and use of
reliable and FIFO channels can be specified using DistAlgo's \co{configure} statement.
For example,
\co{configure clock = Lamport} specifies that Lamport's logical clocks are used; it configures sending and receiving of a message to update the clock value, and defines a function \co{logical\_time()} that returns the clock value.


%% file: inc.tex

\section{Creating Variants using Incrementalization}
\label{sec:inc}



Algorithm variants differ from each other due to different high-level
invariants they maintain and different ways of maintaining them.
We describe the ideas of transforming expensive queries into high-level invariants and using systematic incrementalization to generate efficient algorithms that maintain the query results incrementally.
Each resulting combination of ways of maintaining the invariants forms an algorithm variant.

\paragraph{Example.}

We use as an example Lamport's algorithm for distributed mutual exclusion,
described in his seminal paper that proposed logical clocks%
~\cite{Lam78}.
The problem is for multiple processes to access a shared resource mutually
exclusively, in what is called a critical section, i.e., there can be at
most one process in a critical section at a time.  

Each process can be expressed in DistAlgo as in
Figure~\ref{fig-lam-orig}~\cite{Liu+17DistPL-TOPLAS}.
\begin{figure}[htbp]
\begin{code}
\Sp{} 1 class P extends process:
\Sp{} 2   def setup(s):
\Sp{} 3     self.s := s       # set of all other processes
\Sp{} 4     self.q := \{\}      # set of pending requests\Vex{1}
\Sp{} 5   def mutex(task):
\Sp{} 6     self.t := logical_time()                            # request
\Sp{} 7     send ('request', t, self) to s                      #
\Sp{} 8     q.add(('request', t, self))                         #
\Sp{} 9     await each ('request', t2, p2) in q
\Sp{}                  | (t2,p2) != (t,self) implies (t,self) < (t2,p2)
\Sp{}10           and each p2 in s | some ('ack', t2, =p2) in received | t2 > t
\Sp{}11     task()
\Sp{}12     q.del(('request', t, self))                         # release
\Sp{}13     send ('release', logical_time(), self) to s         #\Vex{1}
\Sp{}14   receive ('request', t2, p2):                          # receive request
\Sp{}15     q.add(('request', t2, p2))                          # 
\Sp{}16     send ('ack', logical_time(), self) to p2            #\Vex{1}                        
\Sp{}17   receive ('release', _, p2):                           # receive release
\Sp{}18     for ('request', t2, =p2) in q:                      # 
\Sp{}19       q.del(('request', t2, p2))                        #
\end{code}\Vex{-2}
  \caption{Lamport's algorithm (lines 6-19) plus setup in DistAlgo.}
  \label{fig-lam-orig}
\end{figure}
The process is set up with sets \co{s} and \co{q} (lines 3-4).
To run a task mutually exclusively, the process sends a request
and adds it to \co{q} (lines 6-8), waits for
(i) own request \co{(t,self)} to be before each other request \co{(t2,p2)}
in \co{q} and (ii) having received an ack with a time \co{t2} later than
\co{t} from each process \co{p2} in \co{s} (lines 9-10) before doing the
task in critical section (line 11), and then removes the request from
\co{q} and sends a release (lines 12-13).  When receiving a request or
release, it sends back an ack and adds to or removes from \co{q} (lines
14-19).

The two conditions in \co{await} are key to the algorithm to ensure mutual
exclusion, while the rest does basic sending and receiving of messages and
bookkeeping of \co{q}.

\paragraph{Incrementalization.}

Incrementalization transforms queries and updates to maintain high-level invariants, including invariants for intermediate and auxiliary values, incrementally~\cite{PaiKoe82,Liu+05OptOOP-OOPSLA,Liu13book,Liu+17DistPL-TOPLAS}.
It can yield diverse algorithms. 

For the example in Figure~\ref{fig-lam-orig}, the two conditions in
\co{await} are queries, consisting of three quantifications including two that are nested; 
and assignments and bookkeeping for \co{s} and \co{q} and implicitly adding to
\co{received} at \co{receive} handlers are updates.

The most direct algorithm can compute queries using iterations, in
\co{for}-loops, whereas an incremental algorithm can maintain the results
of queries at updates and look up the results as needed.  An incrementalized
algorithm maintains high-level invariants not only for the query results but also
for intermediate and auxiliary values needed.  Alternative invariants can often
be used, yielding even greater diversity.

For example, the condition on line 10 in Figure~\ref{fig-lam-orig} can be
transformed into
\begin{code}
    count \{p2: p2 in s, ('ack', t2, p2) in received, t2 > t\} = count s
\end{code}
and then---with variables
\co{responded}, \co{number}, and \co{total} holding the set value and two \co{count} values, respectively, forming three invariants---%
transformed into:
\begin{code}
    number = total
\end{code}
Variable \co{total} is computed at set up of \co{s}, and \co{responded}
and \co{number} at request, and the following \co{receive} handler is
added:
\begin{code}
    receive ('ack', t2, p2):         # new message handler
      if t2 > t:                     # comparison in the condition
        if p2 in s:                  # membership in the condition
          if p2 not in responded:    # test before adding
            responded.add(p2)        # add to responded
            number +:= 1             # increment number
\end{code}
The resulting algorithm differs significantly from direct iteration for the nested quantifications.
The condition on line 10 could also be transformed into two nested \co{count}
queries, and the condition on line 9 can be transformed into a
\co{count} query also, or an aggregate query using \co{min},
yielding different algorithms for incremental maintenance. Details of these
transformations are in~\cite{Liu+17DistPL-TOPLAS}.

In general, incrementalization can also transform nested loops that compute
aggregate values such as sum and min~\cite{Liu+05Array-TOPLAS,GauRaj06,Liu13book}.
For recursive functions as queries, the resulting incremental algorithm can
still use recursion, forming an optimized recursive algorithm, or use
iteration, forming an optimized iterative
algorithm~\cite{liu2000recursion,Liu13book}.
Additionally, more refined data structures can be used to implement sets more efficiently~\cite{Cai+91,Goy00thesis,Liu13book}, such as using one bit for each process in the set \co{responded} above.
Incrementalization also enables new additional optimizations that are made
possible as the results of systematic transformations~\cite{
  Liu13book}.

%% file: sync.tex

\section{Synchronized Execution for DistAlgo}
\label{sec:sync}

A {\em diversified process} is a process with variants.  A system may contain a mixture of diversified and un-diversified processes.  A {\em gateway} process is created for each diversified process.  It represents the variants of a diversified process to the rest of the system, making them appear as a single process.  The gateway intercepts and forwards all inbound and outbound messages of all variants of the diversified process.  We focus on synchronization of DistAlgo constructs; other I/O events, such as file accesses, can be synchronized using standard techniques.

Our synchronized execution framework consists of two parts: (1) an automated program transformation that (1a) ensures all messages are routed via the gateway, and (1b) inserts synchronization with the gateway at yield points, to ensure that all variants have yielded the same number of times before handling their copies of a given inbound message, despite differences in message latency and process execution speed; and (2) an algorithm run by the gateway that determines when to forward messages and when to report divergence (i.e., differences in behavior).  When divergence is reported, the system may initiate application-specific defensive action.


\fullonly{For a simple example of the need for (1b), consider a diversified process whose variants have a receive handler that increments \co{nRcvd} when a message is received and have a \co{run} method containing \co{await nRcvd $\ge$ 1; send nRcvd to p}.  The synchronization inserted at yield points ensures that all variants handle the same number of messages at the yield point associated with this \co{await} statement and hence send the same value.}

We first present the core version of this approach, which assumes all variants of a process have the same communication pattern, i.e., send the same messages to the same destinations in the same order; we discuss later how to relax this assumption. 

\paragraph{Handling Outbound Messages.}

To route outbound messages via the gateway, the transformation replaces all calls to DistAlgo's \co{send} method with calls to \co{send\_sync}, and it inserts a definition of that method in every process class.  \co{send\_sync} sends the original message and its original destination to the gateway.  Processes often send their own process id in messages.  Since each variant of a diversified process has a unique process id, such messages will differ.  To accommodate this as normal behavior, not divergence, \co{send\_sync} replaces all occurrences of the variant's process id in the message with the gateway's process id.  This also reflects the principle that the gateway represents the variants to the rest of the system.  Pragmatically, it ensures that, if the recipient sends a reply to the process id contained in the message, the reply goes to the gateway, as desired.



The gateway stores un-forwarded outbound messages received from each variant in a separate FIFO queue. When all of the queues are non-empty, it compares the messages (including their destinations) at the heads of the queues.  If they are identical, the gateway dequeues the message from all queues and forwards one copy to the destination, otherwise it reports divergence.  To ensure liveness if some divergent variant fails to send a message, once one queue becomes non-empty, the gateway waits a limited amount of time for all queues to become non-empty; if this time limit is exceeded, the gateway reports divergence.

\paragraph{Synchronization at Yield Points and \co{await} Statements.}

The transformation inserts a call to \co{yield\_sync(\p{block}, \p{timeout})} at every yield point, and it inserts a definition of that method in every process class.  The first argument \p{block} is a boolean that indicates whether the yield point is associated with an \co{await} statement.  The second argument \p{timeout}, meaningful when the first argument is \co{True}, is a timeout duration if the \co{timeout} clause is present in that \co{await} statement and is \co{None} otherwise.  The transformation also extends the \co{setup} method to initialize a  variable \co{num\_yields} to zero.  \co{yield\_sync} increments \co{num\_yields}, sends a yield message containing \p{block}, \p{timeout}, and \co{num\_yields} to the gateway and waits for a yield-reply message from the gateway before returning.




The transformation for an \co{await} statement with timeout ensures the total wait time is preserved, even though the waiting period may be split by interactions with the gateway.  
It transforms \co{await $c_1$: $s_1$ or $\ldots$ or $c_k: s_k$ timeout $t$: $s$} into

\begin{center}
\begin{tabular}{@{}r@{~~~~}l@{}}
1 &\co{start\_time = time.time()}\\
2 &\co{while not ($c_1$ or ... or $c_k$):}\\
3 &~~~~\co{elapsed = time.time() - start\_time}\\
4 &~~~~\co{remaining = \p{t} - elapsed}\\
5 &~~~~\co{if remaining $\le$ 0:}\\
6 &~~~~~~~~\co{break}\\
7 &~~~~\co{yield\_sync(True, remaining)}\\
8 &\co{if $c_1$: $s_1$}\\
9 &\co{elif $c_2$: $s_2$}\\
10&$\ldots$\\
11&\co{elif $c_k$: $s_k$}\\
12&\co{else $s$}
\end{tabular}
\end{center}
If the \co{await} statement has no timeout, then lines 1, 3--6, and 12 are omitted, and the second argument of \co{yield\_sync} is \co{None}.

\paragraph{Handling Inbound Messages.}

When the gateway receives an inbound message $m$, it stores $m$ in a queue of un-forwarded inbound messages, waits until it has received yield messages with the same \co{num\_yields} from all variants, forwards to all variants and dequeues all un-forwarded inbound messages, and then sends a yield-reply message to all variants.  The gateway communicates with the variants over FIFO channels, so all variants handle the forwarded messages before proceeding from the current yield point.  In the copy of $m$ to be forwarded to variant $p$, the gateway replaces all occurrences of its own process id with $p$'s process id.

If the gateway has received a yield message from all variants, and has no inbound message to forward to them, its behavior depends on the values of \p{block} and \p{timeout} in the yield messages (if the values of \p{block} differ, or the values of \p{timeout} differ by more than a small amount, divergence is reported).  If \p{block}=\co{False}, the gateway sends a yield-reply message to all variants, allowing them to proceed.  If \p{block}=\co{True} and \p{timeout}=\co{None}, the gateway waits until it has received and forwarded an inbound message before sending a yield-reply message, since the conditions in the \co{await} statements will remain false until the variants' states are updated by handling of an inbound message.  If \p{block}=\co{True} and \p{timeout} is a number, the gateway behaves as in the previous sentence, except it will also send a yield-reply message after time \p{timeout} has elapsed.



\paragraph{Process Creation.} 

The program transformation reads a configuration file that specifies which process types are diversified and the types of their variants.  For each diversified process type $P$, a gateway type \co{Gateway$P$} is generated (basically, this is done by instantiating template code
with the type $P$ and the types of its variants), and process creation statements with type $P$ are transformed to create instances of \co{Gateway$P$} instead.  The \co{setup} method of \co{Gateway$P$} creates an instance of each of the specified variant types, and passes the gateway's process id to the variants as an additional argument to their \co{setup} methods, which are transformed to accept this additional argument.

\paragraph{Relaxed Synchronization.}

The above approach effectively introduces a barrier synchronization for a diversified process's variants at each synchronization point.  This ensures the most timely detection of divergence.  An alternative approach, used in some other synchronized execution frameworks \cite{varan2015,bunshin2017}, is to allow one variant (the ``leader'') to get ahead, try to make the actions of the other processes (the ``followers'') consistent with the leader's actions (e.g., by delivering the same number of messages at the corresponding yield event), and reporting divergence when this is not possible.  This may provide speedup but allows a divergent leader to perform divergent actions before the leader's divergence is detected; when this is unacceptable, such actions should not be allowed to have externally visible effects until the followers catch up and agree on the actions.

\paragraph{Allowing Differences in Message Pattern.}

It may be desirable to relax the requirement that corresponding messages sent by all variants of a process are identical, in order to allow greater diversity.  For example, Lamport's distributed mutual exclusion algorithm \cite{Lam78} sends in \co{ack} messages the current value of the sender's logical clock, whereas the variant in \cite[Fig. 3]{LogicalClock18-ApPLIED} sends in \co{ack} messages the logical time of the request being acknowledged.  To support algorithm variants that have the same communication pattern but different message content, we modify the gateway to omit the equality check on outbound messages when the destination is a diversified process, in which case the gateway sends to the other gateway an array containing the message from each variant, which forwards each message in the array to its corresponding variant.  The correspondence is determined by indexing variants in the order that their types are listed in the configuration file.



To support algorithm variants with different communication patterns, the configuration file can  specify that certain types of messages are {\it un-synchronized}.  When the gateway receives a message of an un-synchronized type from its $i$'th variant, it immediately forwards the message to the destination's gateway, which forwards the message to its $i$'th variant.  For example, for synchronized execution of Lamport's distributed mutual exclusion algorithm and Ricart-Agrawala's distributed mutual exclusion algorithm \cite{RA81}, we specify that \co{ack} and \co{release} messages (used only in Lamport's algorithm) and \co{response} messages (used only in Ricart-Agrawala's algorithm) are un-synchronized; the gateway still synchronizes messages of other types.



%% file: metrics.tex

\section{Diversity Metrics and Runtime Monitoring Tools}
\label{sec:metrics}

\subsection{Code Diversity}

Since diversity is the complement of similarity, we measure code diversity with a well-established document similarity metric, namely, $n$-gram similarity with winnowing \cite{winnowing2003}, which is used in the popular software plagiarism detection tool \href{https://theory.stanford.edu/~aiken/moss/}{Moss} to measure similarity of program source code.  We apply it to Python bytecode, specifically, the sequence of bytecode instructions in a compiled program.
Bytecode similarity is more relevant than source-level similarity, because diversity at the Python level aims to increase resilience to flaws in the runtime system, and bytecode is the program representation used by the runtime system. 
 
An {\em $n$-gram} is a sequence of $n$ consecutive instructions, starting at any position.  The algorithm computes the hash of every $n$-gram in the compiled program, and then (for scalability) selects a subset of those hashes and stores them in a set called the program's {\em fingerprint}.  The number of selected hashes is controlled indirectly by an algorithm parameter $w$ called the {\em window size}.  A window of size $w$ consists of the hashes of $w$ consecutive $n$-grams in the program.  The winnowing algorithm is guaranteed to select at least one hash from each window of size $w$, although it may select more.

A robust metric should have the property that a slightly modified program has high similarity to the original program.  In Python bytecode, local variables and global variables are identified by index.  Inserting one new global variable at the beginning of the program causes renumbering of all global variables; this could make the metric non-robust.  To ensure robustness, we normalize variable indices within each $n$-gram: we re-index the first global variable accessed in the $n$-gram as 0, the second one as 1, etc., and similarly for local variables.  For similar reasons, we replace absolute line numbers used as targets in jump instructions with a place holder.



We quantify code diversity (and similarity) of two programs as 1 minus the Jaccard similarity of their fingerprints.  Recall that the Jaccard similarity of sets $S$ and $T$ is $|S \cap T|/|S \cup T|$.
We 
use 1 minus Jaccard similarity so larger values indicate greater diversity.

An alternative to $n$-gram similarity is Levenshtein distance (a.k.a. edit distance, namely, the minimum number of single-element insertions, deletions, and substitutions needed to change one string to another) between the bytecode sequences in the compiled programs.  Levenshtein distance is less suitable here, because it is sensitive to bytecode orderings in the compiled program that may be unimportant at runtime.  For example, permuting the order in which function definitions appear in the compiled program has no effect on the program's runtime behavior but has a large effect on the Levenshtein distance.  Similarly, swapping the branches in a conditional statement and negating the condition yields an equivalent program with high $n$-gram similarity to the original but (if the branches are large) a large Levenshtein distance from the original.

\subsection{Trace Diversity}

Trace diversity measures the similarity of the sequences of bytecode instructions executed by two programs.  Our bytecode-level tracing tool uses the standard Python \co{trace} module to obtain a source-level trace, and then translates it to a bytecode trace.  A ``blacklist'' of modules to be ignored during the conversion can be specified; in experiments, we blacklist some system modules, such as \co{bootstrap} and \co{trace}.  For each source line mentioned in the trace, identified by filename and line number, the translator compiles that .py file to a .pyc file,
loads the .pyc file using the \co{marshal} module to obtain a \co{code} object, repeatedly uses the \co{dis} (disassembler) module to obtain the bytecode for the entire program as a list of \co{Instruction} objects, and uses the source line number information in the \co{Instruction} objects to determine the sequence of instructions corresponding to each line of source code in that file.  In the traces to be compared, we include only the \co{opcode} and \co{argument} attributes of each \co{Instruction}; other attributes (e.g., \co{is\_jump\_target}) are less important.  We quantify similarity of two traces as the Levenshtein distance (edit distance) between them divided by their average length, for normalization.

\subsection{Input Access Diversity}

Input access diversity measures the similarity of sequences of accesses to input data by two programs, quantified as Levenshtein distance between the sequences divided by their average length, for normalization.  The core of the implementation is a general tool to intercept accesses to attributes of selected objects, by overriding the \co{\_\_getattribute\_\_} method of appropriate classes.  In our use case, the overriding method logs the access and then calls the original \co{\_\_getattribute\_\_} method. For user-defined classes, this is easily accomplished by inserting a definition of \co{\_\_getattribute\_\_} in the class.  This approach does not work for built-in types such as \co{int}, \co{string}, and \co{tuple}, which are common types of input data.  

For each of these built-in classes, we define a new class, e.g. \co{tracked\_int} for \co{int}, that inherits from the built-in class and overrides the \co{\_\_getattribute\_\_} method.  In the remainder of the description, we focus on \co{int}; other built-in types are handled similarly.  The problem is that
some accesses to attributes of \co{tracked\_int} are not logged, because attributes of built-in types are sometimes accessed directly by C code in the CPython runtime system.  For example, even if \co{x} is a \co{tracked\_int}, the addition operator in \co{x+y} compiles to the bytecode instruction \co{BINARY\_ADD}, which does not call \co{\_\_getattribute\_\_} on either argument.  


We overcome this problem by augmenting \co{tracked\_int} to override all methods of \co{int} that access the integer value: \co{\_\_add\_\_}, \co{\_\_eq\_\_}, \co{\_\_le\_\_}, etc.   If \co{x} is a \co{tracked\_int}, an expression like \co{x+y} now compiles to bytecode that uses the \co{CALL\_FUNCTION} instruction to explicitly invoke \co{x}'s \co{\_\_add\_\_} method with argument \co{y}.   The \co{tracked\_int.\_\_add\_\_} method logs the access to the first argument (\co{self}), calls \co{\_\_getattribute\_\_} on the second argument (so the access to it will be logged, if it is a \co{tracked\_int}), and then calls the built-in \co{\_\_add\_\_} method.  Since we need to override these operations anyway, we augment log entries to indicate which operation was performed on the accessed attribute.

If \co{x} is an \co{int}, not a \co{tracked\_int}, 
then \co{CALL\_FUNCTION} invokes the built-in \co{\_\_add\_\_} method, which is implemented by C code that accesses the second argument without calling \co{\_\_getattribute\_\_}.  Consequently, accesses to \co{y} are not logged, even if \co{y} is a \co{tracked\_int}.  To overcome this remaining problem, we modify the program to replace the remaining uses of \co{int} with a new class \co{my\_int}, which inherits from \co{int} and overrides each two-argument method of \co{int} with a method that calls \co{\_\_getattribute\_\_} on the second argument and then calls the original method.  To accomplish this replacement, we bind the name \co{int} to our class \co{my\_int}, using the assignment \co{int = my\_int}.  As a result, a constructor call such as \co{int(1)} returns an instance of \co{my\_int}.  The literal \co{1} still produces an \co{int}.  Therefore, we transform the source program to replace literals with constructor calls, e.g., \co{1} with \co{int(1)}.

The remaining aspects of input access tracking differ for Python and DistAlgo.  These aspects are (1) determining which objects are tracked, and (2) creating meaningful identifiers for tracked objects.  We could easily use the result of Python's built-in \co{id} function to identify objects, but it would be difficult to compare input access traces from different variants (or even different runs of the same variant), because the object identifiers in them would be unrelated.  Instead, we create object identifiers that can be compared meaningfully with object identifiers in other logs, as described below.  The identifier is stored in an attribute of each tracked object.

\paragraph{Python.}

For Python programs, the user specifies which objects should be tracked by modifying the program to make them instances of tracked classes.  For convenience, our
\co{tracker} class provides a method that recursively traverses an object or collection (dictionary, list, tuple, or set) and replaces all instances of trackable built-in types (i.e., types for which a corresponding tracked type exists) with instances of tracked types.  In our benchmark programs, inserting one or two calls to this method suffices.  Tracked objects are identified by a sequence number assigned in the order that the objects are created.
When tracked objects are used for data read as input, these identifiers are meaningful across logs from different variants, because the variants are given the same inputs and hence read the inputs in the same order.

\paragraph{DistAlgo.}

For DistAlgo programs, all messages are automatically considered as inputs; additional inputs, if any, are handled as described above for Python programs.  Instances of trackable built-in types in messages are automatically replaced with instances of tracked types.  Our \co{tracker} class, which inherits from DistAlgo's \co{process} class, is automatically inserted as a parent class of every process class in the given program.  It overrides \co{process.send} with a method that replaces all instances of trackable built-in types in the message with instances of tracked types.

To create meaningful identifiers for tracked objects received in messages, we observe that such an identifier should identify the message in which the object was received.  Our identifier for such an object is a tuple ({\it host}, {\it procNum}, {\it msgNum}, {\it objNum}), where {\it host} is the host on which the sender is running, {\it procNum} identifies the sending process relative to the host, {\it msgNum} identifies the message relative to the sending process, and {\it objNum} identifies the object within the message.  

To avoid dependence on standard process identifiers that cannot be meaningfully compared across executions, we identify processes by a sequence number assigned in the order in which the processes are created.  The \co{tracker} class overrides \co{process.setup} with a method that assigns the process sequence number; \co{tracker.setup} stores the next available process sequence number in a local file.   {\it msgNum} is a per-sender sequence number assigned in the order in which messages are sent.  The object sequence number {\it objNum} is assigned to each object in the message in the order that the object is encountered in a depth-first traversal of the message.  


Input access logs for DistAlgo programs also contain entries corresponding to \co{receive} events, so we can determine that a particular data item (possibly received in a previous message and stored in a data structure) was accessed while processing a particular message.




%% file: evaluation.tex

\section{Evaluation}
\label{sec:eval}

We evaluated our approach on several sequential and distributed algorithms, using Python 3.7.2 and DistAlgo 1.1.0b13.  Our software is available at \url{https://www.cs.stonybrook.edu/~stoller/software}.

For each problem and each diversity metric, we measure the diversity achieved (1) by algorithm diversity alone by averaging the diversity metric for each pair of algorithms; (2) by implementation-level diversity (ILD) alone by averaging the diversity metric for each pair of an algorithm and its ILD variant (i.e., the variant obtained by applying ILD transformations to it); (3) by both forms of diversity together by averaging the diversity metric for each pair of an algorithm and the ILD variant of another algorithm.  For code diversity, we used $n=5$ (the value used in \cite{winnowing2003}), and we disabled winnowing (i.e., included all hashes in the fingerprint), because the bytecode for our examples is not too large.  Library code is not included in our code diversity measurements.

\paragraph{Implementation-Level Diversity (ILD).}

We created ILD by applying these typical ILD transformations: (1) NOP insertion: after each line of code, insert a \co{pass} statement with probability 0.05; (2) instruction reordering: for each two adjacent independent lines of code, swap them with probability 0.5; (3) branch reordering: for each if-statement, swap the branches and negate the condition (if there is no \co{else} branch, pretend \co{else: pass} is present) with probability 0.5; (4) function (including \co{receive} handler) reordering: for each two adjacent independent \co{def} statements, swap them with probability 0.5; (5) argument reordering: for each function (excluding \co{run}, \co{setup}, and receive handlers), swap the first two arguments, swap the third and fourth arguments, etc.; (6) field reordering: reorder the assignment statements that initialize the fields in each class, by swapping the first two, the third and fourth, etc.  Applying more complicated implementation-level diversity techniques is future work; it will require significant effort, because existing implementations of those techniques do not handle Python. 

\subsection{Sequential Algorithms}

Our experiments use these algorithms for these problems: (1) graph reachability: original (iterative) algorithm, incrementalized algorithm, and rule-based algorithm (generated from rules using the method in \cite{liu2009rules}); (2) Hanoi Tower: original recursive algorithm, optimized recursive algorithm, optimized iterative algorithm, and optimized iterative algorithm with swap; (3) longest common subsequence (LCS): original recursive algorithm, optimized recursive algorithm, and optimized iterative algorithm; (4) pattern searching: naive algorithm, Knuth Morris Pratt (KMP) algorithm, Rabin Karp algorithm; (5) sorting: heap sort, quicksort, insertion sort, and merge sort; (6) tree search: recursive and iterative algorithms for AVL trees, recursive algorithm for B-trees, iterative algorithm for red-black trees, and recursive and iterative algorithms for (unbalanced) binary search trees. 

The results are in Table \ref{tab:seq}.  We see from the last column that, for all three metrics, algorithm diversity creates more diversity than ILD, and that the two together create even more.

\begin{table}[tb]
\centering
\begin{tabular}{@{}l|l||c|c|c|c|c|c|c@{}}
    \hline
    Metric   & Level\, & reach. &  Hanoi & LCS  & pat.~search. & sort & tree search & \,Avg.\\
             &         & 3 variants & 4 variants & 3 variants & 3 variants & 4 variants & 6 variants&\\
    \hline\hline
    Code     & algo    & 0.80  & 0.58  & 0.65 & 0.81  & 0.79  & 0.83 & 0.74\\
    Diversity& impl    & 0.40  & 0.39  & 0.66 & 0.52  & 0.32  & 0.63 & 0.49\\
             & both    & 0.80  & 0.65  & 0.82 & 0.83  & 0.80  & 0.89 & 0.80\\
    \hline
    Input Access& algo  & 1.04  & 0.54  & 0.58 & 0.28  & 0.77  & 0.35 & 0.59\\
    Diversity& impl    & 0     & 0.18  & 0.82 & 0.21  & 0     & 0    & 0.20\\
             & both    & 1.04  & 0.57  & 1.12 & 0.28  & 0.77  & 0.36 & 0.69\\
    \hline
    Trace    & algo    & 1.45  & 0.42  & 1.22 & 0.69  & 0.81  & 0.80 & 0.90\\
    Diversity& impl    & 0.05  & 0.30  & 0.60 & 0.23  & 0.11  & 0.14 & 0.23\\
             & both    & 1.45  & 0.45  & 1.39 & 0.70  & 0.82  & 0.82 & 0.94\\
    \hline
\end{tabular}\Vex{1}
\caption{Experimental results for sequential algorithms.  In the ``Level'' column, ``algo'' and ''impl'' denote algorithm and implementation-level diversity, respectively.  The last column contains averages.}
\label{tab:seq}
\end{table}




\subsection{Distributed Algorithms}

Our experiments use the following algorithms: (1) 2-phase commit (2PC); (2) Hirschberg-Sinclair's leader election (HSleader) \cite{HS80}; (3) Lamport's distributed mutual exclusion (Lamutex)~\cite{Lam78}; (4) Lamport's basic Paxos \cite{Lam01paxos}; (5) Ricart-Agrawala's distributed mutual exclusion (RAmutex)~\cite{RA81}.  We used configurations with 3 or 4 processes for each algorithm.  There are two variants of each algorithm: one variant that uses high-level queries over message histories, and one that explicitly maintains the result of those queries (and related intermediate results and auxiliary values), updating them in assignment statements, especially in \co{receive} handlers. 

When measuring the dynamic metrics, we avoid spurious differences between the variants due to the platform's scheduling variability by running all variants in parallel using synchronized execution (for programs other than 2PC, due to a bug that we are still resolving in the interaction between our program transformations for synchronized execution and input access tracing, when measuring input access diversity, we instead avoided such spurious differences by running the variants separately but each with the same pattern of injected message delays that are larger than the platform's scheduling variability and designed to avoid races in message delivery order).

The results are in Table \ref{tab:dist}.  We see from the last column that, for all three metrics, algorithm diversity creates significantly more diversity than ILD.   The trace diversity produced by ILD is considerably smaller than the input access diversity it creates.  These results are not inconsistent, because both are measured as ratios, and input accesses constitute a small fraction of the program's full activity recorded in the bytecode trace.  The results for trace diversity for algorithm diversity for distributed algorithms are notably smaller than for sequential algorithms, because the trace includes execution of DistAlgo runtime library for networking, which is not diversified.

\begin{table}[tb]
\centering
\begin{tabular}{@{}l@{\,}|@{\,}l@{~}||c|c|c|c|c|c@{}}
    \hline
    Metric   & Level & ~~~~2PC~~~~~  & ~HSleader~ & ~Lamutex~ & ~~~Paxos~~~ & \,RAmutex\, & \,Average\\
    \hline\hline
    Code     & algo   &0.56   &0.66    &0.50    &0.68    &0.53   &0.59\\
    Diversity& impl   &0.19   &0.18    &0.08    &0.30    &0.27   &0.21\\
             & both   &0.59   &0.68    &0.53    &0.68    &0.54   &0.60\\
    \hline
    Input Access&algo & 1.10  & 0.47   & 0.21   & 0.28   & 0.61  & 0.53 \\
    Diversity& impl   & 0.08  & 0.04   & 0      & 0.03   & 0.17  & 0.06\\
             & both   & 1.09  & 0.52   & 0.21   & 0.30   & 0.61  & 0.55\\
    \hline
    Trace    & algo   & 0.20  & 0.35   & 0.13   & 0.54   & 0.21  & 0.29\\
    Diversity& impl   & 0.06  & 0.03   & 0.02   & 0.13   & 0.04  & 0.06\\
             & both   & 0.20  & 0.36   & 0.14   & 0.52   & 0.21  & 0.29\\
    \hline
\end{tabular}\Vex{1}
\caption{Experimental results for distributed algorithms, with 2 variants for each algorithm. In the ``Level'' column, ``algo'' and ''impl'' denote algorithm diversity and implementation-level diversity, respectively. }
\label{tab:dist}
\end{table}






%% file: related.tex

\section{Related Work}
\label{sec:related}

Existing techniques for automated software diversity, including all those surveyed in \cite{larsen2014}, create implementation-level diversity, changing details of the implementation without changing the algorithm.  Typically this is done by applying relatively simple local transformations, like those used in our evaluation.  There are also some complex global transformations, such as instruction set randomization.  These transformations are fully automated and more easily applied to large programs, but they are limited in that they do not create algorithm diversity.  For example, they do not change the pattern in which inputs are used by the program.


Most work on automated software diversity for resilience transforms C programs or (disassembled) machine code, for broader applicability to systems code.  There is some work on automated diversity for programs in JIT-compiled high-level languages, which diversifies the machine code generated by the JIT compiler.  For example, librando does this for Java and JavaScript \cite{librando2013}, and INSeRT does this for JavaScript \cite{insert2011}.  This low-level approach is suitable for creating implementation-level diversity.  Our methodology diversifies the high-level program directly to create algorithm diversity.

In $N$-version programming  \cite{Nversion1985}, $N$ versions of a system (or component) are created by separate and independent manual design and implementation efforts starting from the same requirements specification, and the versions are run in parallel with synchronized execution.  The goal is resilience in the presence of design faults, since independent teams are less likely to make the same design mistakes.  Our work, like other work on software diversity, aims to mitigate software vulnerabilities, not design errors.  The two techniques could be used together to address both.  $N$-version programming may introduce algorithm diversity, but not in a controlled way, and at the cost of significant manual effort.  In contrast, our approach is to create variants using a program transformation and optimization method based on  systematic incrementalization, which guides the process, helps control how much diversity is introduced, and helps ensure correctness compared to ad-hoc development of variants.  Our program transformation system InvTS \cite{Liu+09Inv-GPCE,Gor+12Compose-PEPM} provides semi-automated support for the method, significantly reducing manual effort. 

Synchronized execution has been widely studied in the fault-tolerance community, where it is often called {\em $N$-version execution}.  $N$-version execution frameworks typically work at the system-call level, so they can be applied to software running on a given operating system, regardless of the application programming language.  Our synchronized execution framework is applicable only to applications written in DistAlgo, but it is more portable and lighter weight.  It can be used on any OS supported by 
DistAlgo (Windows, macOS, Linux, and Android), while system-call based approaches are highly OS-specific, e.g., Varan \cite{varan2015} and Bunshin \cite{bunshin2017} are $N$-version execution frameworks for Ubuntu.  It is lighter-weight because a single high-level synchronization event is typically implemented by multiple system calls. 

\subsection{Evaluation of Diversity Techniques}

A few approaches are commonly used to evaluate implementation-level diversity techniques.  One is to estimate the probability of a successful memory-related exploit (e.g., buffer overflow or format string attack) based on the information about the diversified program that the attacker would need to guess, more specifically, the type of information (e.g., the address of a specific object, or the difference between the addresses of two specific objects) and the number of possible values of that type of information due to the randomization in the diversity transformation.  This approach is used in, e.g., \cite{bhatkar2005,giuffrida2012}.

Diversity techniques designed specifically to defend against ROP attacks are typically evaluated using a {\em coverage metric} that measures the fraction of ROP gadgets re-located by the transformation, and sometimes also an {\em entropy metric} that measures the number of possible new positions of the ROP gadgets, reflecting the probability of the attacker correctly guessing the new locations.  This approach is used in, e.g., \cite{smashing2012,librando2013}).

These approaches based on specific vulnerabilities in low-level languages are  unsuitable for evaluating diversity for interpreted languages, such as Java and Python.

\fullonly{Kil et al.'s address space layout permutation (ASLP) extends ALSLR with address randomization within the code and data segments\cite{ASLP2006}.  They measure the resulting diversity by comparing the addresses of variables in the randomized program with the addresses of the corresponding variables in the original program.  This approach is suitable for low-level languages that allow direct access to the addresses of variables, but not suitable for high-level languages such as Java and Python.}
